\documentclass[aps,prx,twocolumn,a4paper,showkeys,10pt,longbibliography]{revtex4-1}
\usepackage[english]{babel}
\usepackage[T1]{fontenc}

\usepackage[a4paper,top=3cm,bottom=2cm,left=2cm,right=2cm,marginparwidth=1.75cm]{geometry}

\usepackage{amsmath}
\usepackage{amsfonts}
\usepackage{amssymb}
\usepackage{amsthm}
\usepackage{hyperref}
\usepackage{tikz}
\tikzset{>=latex}

\newcommand{\ba}{\begin{align}}
\newcommand{\ea}{\end{align}}
\newcommand{\be}{\begin{equation}}
\newcommand{\ee}{\end{equation}}

\newcommand{\ket}[1]{| #1 \rangle}

\newcommand{\braOket}[3]{\langle \: #1 \: | \: #2 \:| \: #3 \: \rangle}

\makeatletter
\newcommand*{\rom}[1]{\expandafter\@slowromancap\romannumeral #1@}
\makeatother
\newcommand{\kbf}{\mathbf{k}}

\definecolor{myYellow}{rgb}{0.70196,0.34510,0.02353}
\definecolor{myGreen}{rgb}{0.247059,0.619608,0.098039}
\definecolor{myRed}{rgb}{0.647059,0.113726,0.074510}
\definecolor{myBlue}{rgb}{0.02353,0.38039,0.70196}
\definecolor{myPurple}{rgb}{0.3294118,0.15294118,0.533333}
\definecolor{myOrange}{rgb}{0.70196078,0.345098039,0.02352941}

\begin{document}
\title{Multiphoton interaction phase shifts in attosecond science}
\author{Mattias \surname{Bertolino}}
\affiliation{Department of Physics, Lund University, Box 118, SE-221 00 Lund, Sweden}
\author{Jan~Marcus \surname{Dahlstr\"om}}
\affiliation{Department of Physics, Lund University, Box 118, SE-221 00 Lund, Sweden}
\date{\today}

\begin{abstract}
  {\it Ab initio} simulations of a range of interferometric experiments are used to identify a strong dependence on multiphoton phase shifts in above-threshold ionization.
  A simple rule of thumb for interaction phase shifts is derived to explain both the conservation of  photoelectron yield and its {\it absolute} CEP-dependence.
  For instance, it is found that interferometric above-threshold ionization experiments are shifted by $\pi/4$ relative to RABBIT experiments, and that there is no RABBIT-term in a laser-assisted photoionization experiment with odd and even harmonics.
  Thus, our work helps to resolve the issues of interpretation of quantum dynamics in attosecond and free-electron laser sciences.
\end{abstract}

\maketitle

Interference effects have long played a central role for our understanding of both classical and quantum physics.
While traditional spectroscopy allows for measurements of phase differences between separate sources, e.g.\ high-order harmonic generation (HHG) from atoms~\cite{ZernePRA1997} or molecules~\cite{XibinPRL2008,SmirnovaNature2009}, the technique is limited to probe transitions
with the {\it same} photon energy.
In contrast, novel types of nonlinear spectroscopies make use of probe laser beams to change the dynamics in the interferometric arms, thus making it possible to interfere {\it different} energy components of the system~\cite{Kim2014NaturePhotonics}.
Prime examples of such nonlinear arms include perturbed quasi-classical trajectories in HHG in atoms~\cite{dudovichNaturePhysics2006,DahlstromJPB2011,ShafirNature2012,PedatzurNaturePhysics2015}, molecules~\cite{UzanNaturePhotonics2020} and solids~\cite{VampaJPB2017}; perturbed photoelectrons in laser-assisted photoionization (LAP) from atoms~\cite{PaulScience2001,KlunderPRL2011,LaurentPRL2012,DahlstromCP2013,PazourekRevModPhys2015} and molecules~\cite{CaillatPRL2011}; resonant two-photon ionization via bound~\cite{SwobodaPRL2010} or autoionizing states~\cite{JimenezPRL2014,KoturNatCom2016,BarreauPRL2019}; and perturbed photoelectrons in above-threshold ionization (ATI)~\cite{zippOptica2014}.
Nonlinear interferometric arms have recently also found applications in free-electron laser (FEL) experiments~\cite{MarojuNature2020,you_new_2020}.
This plethora of interferometric arms makes it possible to inherently study nonlinear processes, such as HHG and ATI, but they also raise the question of interpretation between unperturbed and perturbed dynamics.

In this Letter we address the interpretation of quantum dynamics in attosecond experiments by performing a systematic study using the time-dependent configuration interaction singles~\cite{rohringer_configuration-interaction-based_2006, GreenmanPRA2010} (TDCIS) method.
Photoelectron distributions are obtained with the time-dependent surface flux~\cite{tao_photo-electron_2012} (t-SURFF) method and compared with  simulations based on the Keldysh-Faisal-Reiss~\cite{keldysh_ionization_1965,faisal_multiple_1973,reiss_effect_1980} (KFR) theory for strong-field ionization.
Specifically, we consider three recent attosecond experiments, which utilize different nonlinear interferometric arms, and compare them with the traditional RABBIT experiment~\cite{PaulScience2001}.
First, we consider a LAP experiment by Laurent et al.~\cite{LaurentPRL2012}, where even and odd harmonics from HHG were coupled by a laser photon from the probe beam.
Second, we consider a LAP experiment by Maroju et al.~\cite{MarojuNature2020}, where attosecond pulse structures of FEL beams were coupled to twin sidebands by a mix of one and two exchanged probe photons.
Lastly, we consider the ATI experiment by Zipp et al.~\cite{zippOptica2014}, where an ATI process was probed by photons of half the frequency.
We find that all three recent experiments exhibit interference patterns that depend on multiphoton phase shifts and on the directionality of the photoelectron, phenomena that are not observed in the RABBIT experiment~\cite{PaulScience2001}.
Using KFR, we derive a simple rule of thumb capable of explaining the general phase and amplitude effects in all above mentioned experiments
and provide a physical interpretation based on  an accumulation of interaction phases in the multiphoton processes.
In this work atomic units are used: $\hbar=e=a_0=m_e=4\pi\epsilon_0=1$.

\begin{figure*}[ht]
  \centering
  \includegraphics{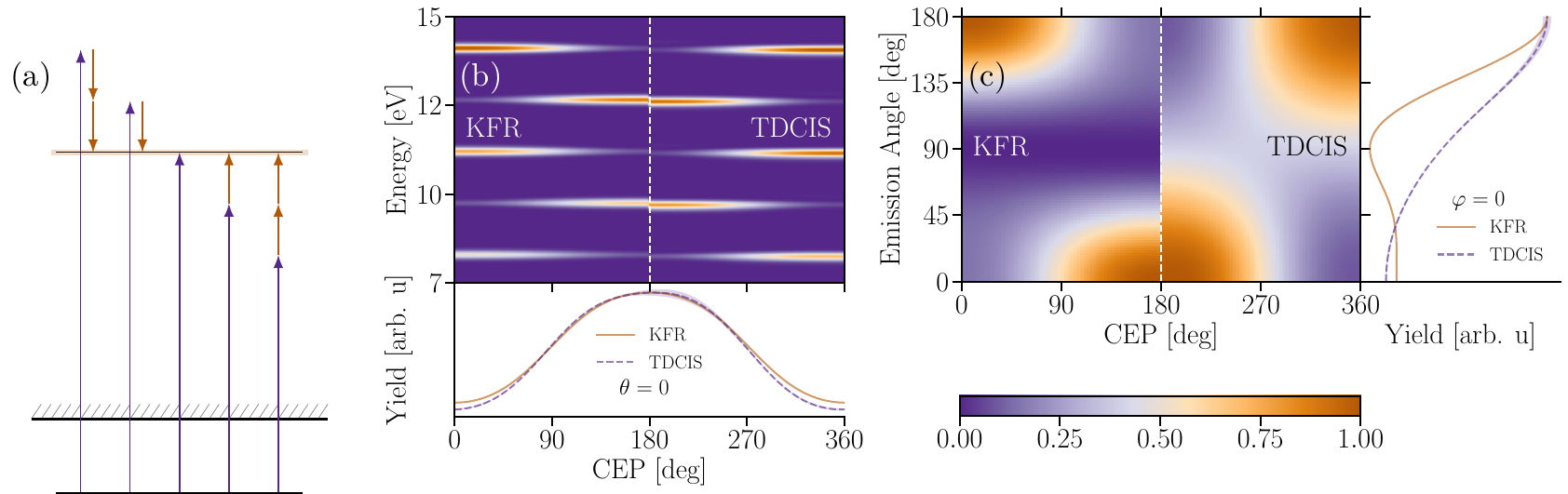}
  \caption{
    (a) Interactions in the rule of thumb for the experiment of Laurent et al.\ in Ref.~\cite{LaurentPRL2012}.
    (b) Photoelectron peaks corresponding to high-order harmonics 20--24 using KFR and TDCIS in Ne2p.
    The relative phase between even and odd harmonics is $\delta=-\pi/2$ and the photoelectrons are measured in the \textit{up} direction.
    The lower part shows the lineout of the $21^{\mathrm{st}}$ harmonic.
    (c) Yield of the $21^{\mathrm{st}}$ harmonic peak resolved in emission angle, $\theta$, and CEP, $\varphi$.
    The right sidepanel shows the lineout of the $21^{\mathrm{st}}$ harmonic over emission angle for $\varphi=0$.
    The TDCIS-lineouts are estimated as an average of calculations with $r_{\mathrm{surf}}=\{84,88,92,96\}$ in t-SURFF.
The purple shaded area marks confidence up to one standard deviation in TDCIS.}
  \label{fig:laurent}
\end{figure*}

In 2012, Laurent et al.\ proposed a phase shift, $\delta=\pi/2$, between odd and even harmonics from HHG, based on an experimental ``checkerboard'' pattern over CEP between the \textit{up} ($\theta=0^{\circ}$) and \textit{down} ($\theta=180^{\circ}$) photoelectrons~\cite{LaurentPRL2012}.
This phase shift has caused controversy because it opposes the idea that one attopulse per cycle is generated from HHG with an $\omega/2\omega$ driving field~\cite{MauritssonPRL2006}.
While high robustness of the phenomena was reported from the experiment, the accompanying analysis was limited to one-photon and two-photon ionization processes~\cite{starace_theory_1982,DahlstromCP2013}, which is insufficient for this type of interferometric process.
 For this reason, we present in Fig.~\ref{fig:laurent}~(b) the result of LAP simulations for {\it up} electrons from a broad comb of odd, $(2n+1)\omega$, and even, $2n\omega$, harmonics coupled by a laser
 probe field, $\omega$.
 We find that the relative phase of $\delta=\pi/2$ leads to a checkerboard structure, while the $\delta=0$ case does not (not shown).
 Thus, we have confirmed the proposal of Laurent et al.~\cite{LaurentPRL2012}, and open the quest for understanding the physical meaning and wide applications of such phase shifts in attosecond experiments.

The simulation presented in the left part of Fig.~\ref{fig:laurent}~(b) is performed using KFR with the initial state $\ket{a(t)}=\ket{a}\exp[i I_p t]$ taken to be a scaled hydrogenic $1s$ orbital with a binding energy equal the Hartree-Fock (HF) $2p$ orbital in neon, $I_p=-\epsilon_{2p}^\mathrm{HF}=23.142$\,eV.
The simulation presented in the right part of Fig.~\ref{fig:laurent}~(b) is performed using TDCIS in velocity gauge with the active space  restricted to excitations from the $2p$ orbital with $m=-1,0,1$, and $I_p=-\epsilon_{2p}^\mathrm{HF}$.
The electric fields are derived from overlapping $\tau=10$\,fs Gaussian vector potentials,
\begin{align}
  A(t) =&
  \left[\Lambda_{\Omega,0}\sum_{i} \sin(\Omega_{i}t - \delta_i) + \Lambda_{\omega,0}\sin(\omega t - \varphi)\right]
  \nonumber \\
  \times&\exp\left[-2\ln(2)\frac{t^2}{\tau^2} \right],
\end{align}
with peak envelopes
$\Lambda_{\Omega,0} = 0.005$ (pump) and
$\Lambda_{\omega,0} =0.04$ (probe) ($1.8\times 10^{11}$\,W/cm$^2$ at photon energy $1.55$\, eV).
In both cases, the photoelectrons exhibit clear $\omega$-modulations over CEP, which shows that neither initial angular-momentum, nor electron correlation affect the phenomena.
In Fig.~\ref{fig:laurent}~(c),
harmonic 21 is shown over both emission angle, $\theta$, and CEP, $\varphi$.
The photoelectron distribution can be parity transformed by a CEP shift of $\Delta\varphi=180^{\circ}$.
While the photoelectron emission along the polarization axis exhibits qualitatively similar modulation over CEP for KFR and TDCIS, the photoelectron angular distributions are different in general, as seen in the sidepanels of Fig.~\ref{fig:laurent}~(b,c) for $\theta=0$ and $\varphi=0$ respectively.
The photoelectron emission distributions exhibit qualitatively similar modulation with CEP for emission along the polarization axis, while qualitatively different angular distributions are found in other emission angles.

In order to interpret the observed CEP phenomena we now consider the general LAP process: $A + \gamma_{\Omega} +n \gamma_{\omega} \to A^+ + e^-$, where $n$ laser photons are exchanged.
The KFR transition amplitude for photoionization can be used to express LAP,
\begin{widetext}
  \begin{align}
  \label{eq:kfr-amplitude}
  c_{\kbf} &\approx \frac{1}{2} \sum_{n=-\infty}^{\infty} (-i)^{n} \exp[in\varphi]
  J_{n} \left(\frac{\kbf \cdot \pmb{\Lambda}_{\omega,0}}{\omega}\right)
  \sum_{m=-\infty}^{\infty} \exp[i2m\varphi]
  J_{m} \left(\frac{U_p}{2\omega}\right) \\
  &\times \braOket{{\kbf}}{\hat{p}_z}{a} \exp[i\delta]
  \int \mathrm{d} t ~\Lambda_{\Omega}(t)
 \nonumber
 \exp[i(\epsilon_k + I_p + U_p - \Omega - n\omega - 2m\omega) t],
\end{align}
\end{widetext}
for an idealized monochromatic probe field using a Jacobi-Anger expansion~\cite{kitzler_quantum_2002}.
The Bessel functions $J_n(k \Lambda_{\omega,0}\cos(\theta)/\omega)$ stem from the $\mathbf{k}\cdot\mathbf{A}(t)$ term and the Bessel functions $J_m(U_p/(2\omega))$ stem from the $A^2(t)/2$ term in the minimal coupling Hamiltonian.
In the limit of a small ponderomotive energy, $U_p = \Lambda_{\omega,0}^2/4 \ll 2\omega$, the sum over $m$ is reduced to the zeroth term because $J_m\approx \delta_{m0}$.
Thus the remaining sum over $n$ can be interpreted as any number of interactions of $\mathbf{k} \cdot \mathbf{A}(t)$ type in the weak field limit appropriate for probe processes.
Bessel functions are odd and even functions depending on the integer $n$.
Although the argument of the oscillatory Bessel functions, $\xi = k \Lambda_{\omega,0}\cos(\theta)/\omega$, is important for the parity of the photoelectron, it is suppressed in the following for brevity.
The LAP process is {\it linear} in the XUV field, which makes it easy to interpret interference features as energy-shifted replicas of the elementary photoionization process, $f_{\kbf}$, for each number of laser interactions, $n$.
We note that the processes of absorption ($n>0$) and emission ($n<0$) of laser photons have the same phase factor because $(-i)^{-n}J_{-n}=(-i)^{n}J_{n}$.
Thus, we have identified an {\it interaction phase} that accumulates with each interaction with the laser field, and the $n^\mathrm{th}$ LAP process can be written as
\begin{equation}
  \label{eq:thumbrule}
  c_{\kbf}^{(n)} \approx
  (-i)^{|n|} 
  J_{|n|}
  \exp(i n \varphi) 
  f_{\kbf}^{(\mathrm{pump})},
\end{equation}
referred to as the ``rule of thumb'' for ATI, valid in the limit of a small ponderomotive energy.

Eq.~(\ref{eq:thumbrule}) is applied for probe-photon processes up to second order, illustrated in Fig.~\ref{fig:laurent}~(a), for a plateau of equally strong harmonics, with a phase shift of $\delta$ between even and odd harmonics: $f_{\kbf}^{(\mathrm{odd})} = 1$ and $f_{\kbf}^{(\mathrm{even})} = \exp[i\delta]$.
The transition amplitude for the laser-assisted harmonics are
\begin{widetext}
\begin{align}
  \begin{cases}
  \label{eq:laurent-ampl}
 c_{\kbf}^{(\mathrm{odd})} &=
  - J_2\exp[-i2\varphi]
  - i J_1\exp[-i(\varphi-\delta)]
  + J_0
  - i J_1 \exp[i(\varphi + \delta)]
  - J_2  \exp[i2\varphi]  \\
 c_{\kbf}^{(\mathrm{even})} &=
  - J_2 \exp[-i(2\varphi - \delta)]
  - i J_1 \exp[-i\varphi]
  + J_0 \exp[i\delta]
  - i J_1 \exp[i\varphi]
  - J_2 \exp[i(2\varphi + \delta)].
   \end{cases}
   \\
   \end{align}
\end{widetext}
This leads to photoelectron peaks that modulate with the CEP as
\begin{equation}
  \label{eq:laurent-prob-odd}
  \begin{cases}
    |c_{\kbf}^{(\mathrm{odd})}|^2 & =1 + 2\xi\sin(\delta)\cos(\varphi) + \mathcal{O}(\xi^3) \\
    |c_{\kbf}^{(\mathrm{even})}|^2 & =1 - 2\xi\sin(\delta)\cos(\varphi) + \mathcal{O}(\xi^3),
  \end{cases}
\end{equation}
where $\mathcal{O}(\xi^3)$ indicates a third-order error in the probe field.
While the CEP-dependent term vanishes for $\delta=0$, it is maximal for $\delta=(\frac{1}{2}+N)\pi$, leading to the checkerboard pattern.
In agreement with our numerical simulations, presented in Fig.~\ref{fig:laurent}~(b), there is no $2\omega$-modulation term in Eq.~(\ref{eq:laurent-prob-odd}).
The ``RABBIT''-term, proposed by Laurent et al.~\cite{LaurentPRL2012}, is cancelled out by  transitions with two probe photons, which accumulate a $\pi$-shift from two interaction phases, relative to the direct path.
Thus, the rule of thumb explains the robustness of the  checkerboard pattern with substantial increase in laser intensity observed experimentally~\cite{LaurentPRL2012} and that the total probability of even and odd pairs is conserved in the plateau.

In RABBIT experiments~\cite{PaulScience2001}, photoionization by a comb of odd harmonics is perturbed by a laser field to generate background-free sideband peaks (SB).
While the sideband modulation over CEP is the main observable in RABBIT, recent work has also included the CEP-modulation of the odd harmonics (HH) to improve the experimental statistics without any formal justification~\cite{IsingerScience2017}.
Given a sequence of synchronized {\it odd} harmonics, $f_{\kbf}^{(\mathrm{odd})}=1$ and $f_{\kbf}^{(\mathrm{even})}=0$, the rule of thumb gives
\begin{equation}
  \label{eq:rabbit-prob}
  \begin{cases}
  |c_{\kbf}^{(\mathrm{SB})}|^2
  &= \xi^2\cos^2(\varphi) + \mathcal{O}(\xi^4) \\
  |c_{\kbf}^{(\mathrm{HH})}|^2
  &= 1 - \xi^2\cos^2(\varphi) + \mathcal{O}(\xi^4).
  \end{cases}
\end{equation}
The fact that SB and HH peaks in RABBIT behave in opposite ways with CEP is a phenomenon directly related to multiphoton interaction phase shifts.
Eq.~(\ref{eq:rabbit-prob}) also shows that the probability of SB+HH pairs are conserved.

\begin{figure*}
  \centering
  \includegraphics{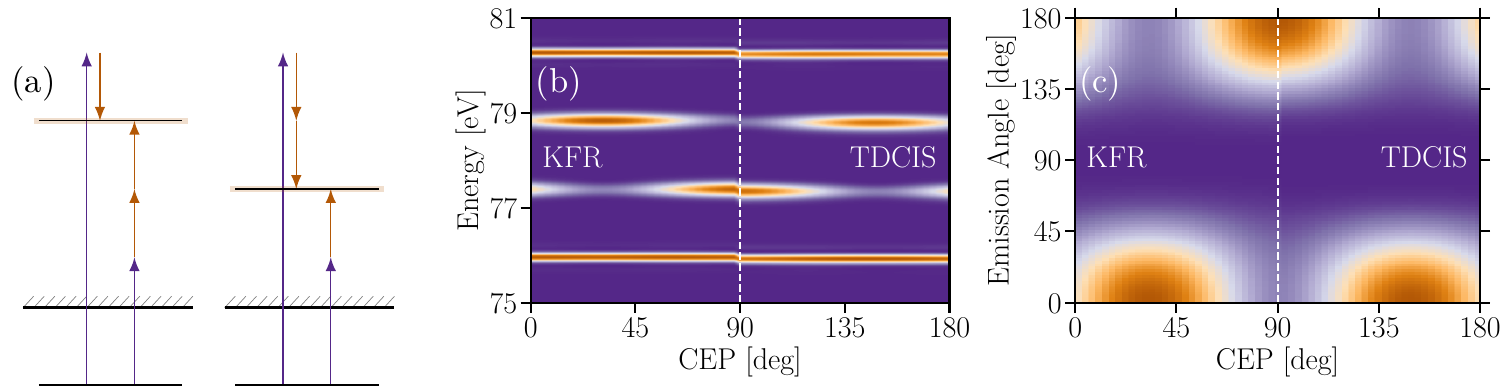}
  \caption{
  (a) Interactions in the rule of thumb for the twin sideband experiment by Maroju et al. in Ref.~\cite{MarojuNature2020}.
  (b) Photoelectron peaks corresponding to FEL-generated harmonics 64 and 67 using KFR and TDCIS in Ne2p.
  The photoelectrons are measured in the \textit{up} direction.
  (c) Yield of the high twin sideband resolved in emission angle, $\theta$, and CEP, $\varphi$.}
  \label{fig:Maroju}
\end{figure*}

In Fig.~\ref{fig:Maroju}~(b), we show the result of simulations of the FEL experiment by Maroju et al.~\cite{MarojuNature2020}, where two FEL beams with a photon energy difference of $\Delta\Omega^{\mathrm{FEL}} = \Omega_>^{\mathrm{FEL}} - \Omega_<^{\mathrm{FEL}} = 3\,\omega$ are used to photoionize neon atoms with an assisting laser with photon energy $\omega$.
This results in the formation of twin sidebands that are created by exchange of a combination of one and two laser photons, as illustrated in Fig.~\ref{fig:Maroju}~(a).
The ``low'' (``high'') twin is generated by absorption of one (two) laser photon and emission of two (one) laser photons.
In contrast to the above studied experiments, the present result shows a non-trivial CEP-translation.
In order to understand the CEP-dependence, we apply the rule of thumb, with $f_{\kbf}^>= f_{\kbf}^<=1,$ which leads to
\begin{align}
\label{eq:Maroju-amp}
\begin{cases}
 c_{\kbf}^{(\textrm{high})} &=
 -i J_{1} \exp[-i\varphi]-J_{2}\exp[ i2\varphi] \\
 c_{\kbf}^{(\textrm{low})} &=
 -J_{2}\exp[-i2\varphi]-i J_{1} \exp[ i\varphi].
\end{cases}
\end{align}
This shows that the two terms in each amplitude are out of phase by $-\pi/2$ at $\varphi=0$, due to the different orders of interactions with the laser field.
The corresponding probability is
\begin{equation}
  \label{eq:Maroju-prob}
  \begin{cases}
    |c_{\kbf}^{(\mathrm{high})}|^2 &
    = \frac{\xi^2}{4} + \frac{\xi^3}{8}\sin(3\varphi) + \mathcal{O}(\xi^4) \\
    |c_{\kbf }^{(\mathrm{low})}|^2 &
    = \frac{\xi^2}{4} - \frac{\xi^3}{8}\sin(3\varphi) + \mathcal{O}(\xi^4),
  \end{cases}
\end{equation}
showing that the two sidebands have equal strength for $\varphi=0$, but then evolve in different ways with CEP.
Further, the relative sign of the two terms in Eq.~(\ref{eq:Maroju-amp}) is flipped under a parity transformation due to the properties of the Bessel functions.
In this sense Maroju's experiment is a {\it higher-order} version of the orbital parity mixing experiment by Laurent et al.~\cite{LaurentPRL2012}.
Indeed, in Fig.~\ref{fig:Maroju}~(c), it is shown that it is possible to perform a parity transformation of the twin peaks by tuning the CEP of the laser field by $\Delta \varphi=60$ degrees.

\begin{figure*}
  \centering
  \includegraphics{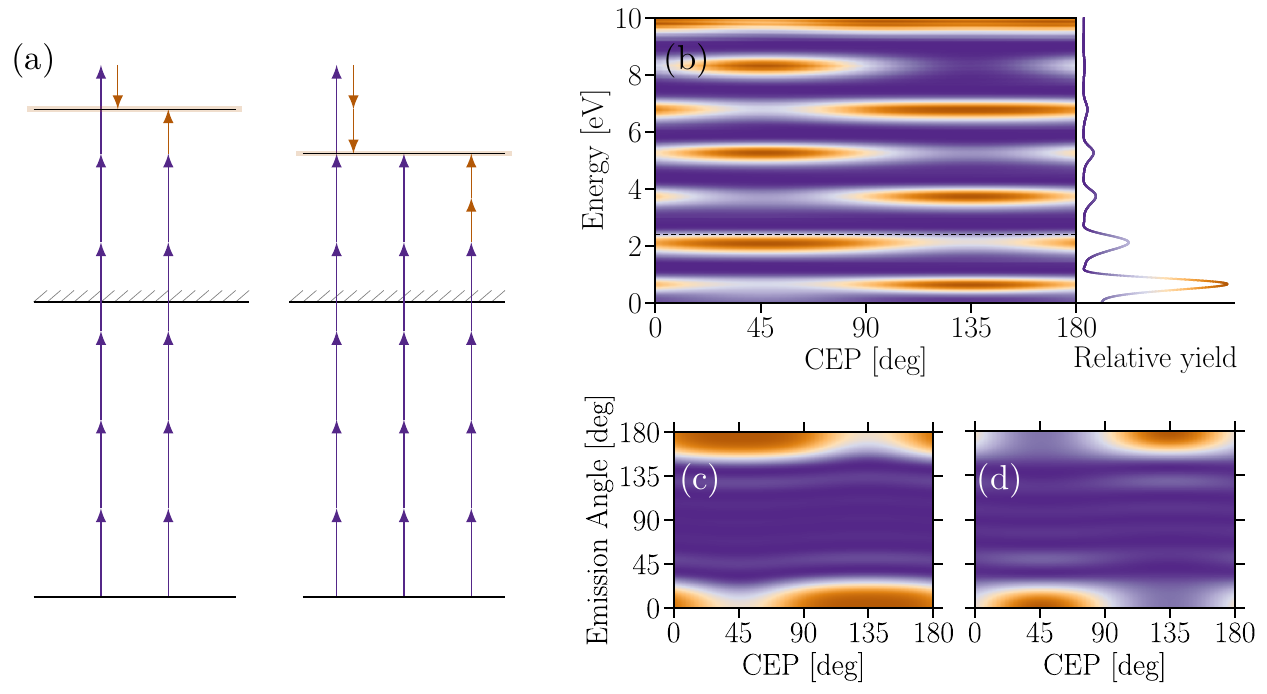}
  \caption{
  (a) Interactions in the rule of thumb for the $2\omega/\omega$-ATI experiment by Zipp et al. in Ref.~\cite{zippOptica2014}.
  (b) Photoelectron sideband and ATI peaks in Ne2p measured in the upper $z$-hemisphere using KFR.
  $2 U_p$ is indicated by the black dashed line.
  Yield of the (c) third sideband [6.8 eV in (b)] and (d) second ATI peak [5.3 eV in (b)] with resolved in emission angle, $\theta$, and CEP, $\varphi$.}
  \label{fig:zipp}
\end{figure*}

Finally, we present in Fig.~\ref{fig:zipp}~(b) ATI simulations for the experiment by Zipp et al.~\cite{zippOptica2014}.
The analysis is limited to KFR, which neglects the rescattering process that contributes to photoelectrons with kinetic energy above $2U_p$, for simplicity.
The initial wave packet is here created with a strong $2\omega$-field ($8\times 10^{13}$\, W/cm$^{2}$) by the nonlinear ATI-process,
rather than the linear photoionization process.
Intermediate sidebands (SB) are created by the exchange of $\omega$ photons from a weaker laser field ($4\times 10^{11}$\, W/cm$^{2}$), as illustrated in the left diagram of Fig.~\ref{fig:zipp}~(a).
The ATI peaks are further perturbed by the $\omega$ laser field as illustrated in the right diagram of Fig.~\ref{fig:zipp}~(a).
The data presented in  Fig.~\ref{fig:zipp}~(b) is normalized to each ATI and SB peak individually.
In reality the probability of ATI decreases with increasing order, see the insert in Fig.~\ref{fig:zipp}~(b).
While there is a great level of similarity between the $2\omega/\omega$-ATI simulation in Fig.~\ref{fig:zipp}~(b) and the RABBIT modulations in Eq.~(\ref{eq:rabbit-prob}), there is a marked difference in the CEP-dependence on the absolute scale by $\Delta\varphi \sim \frac{\pi}{4}$.
To understand this numerical result, we turn to the rule of thumb.
In ATI, interaction phases will accumulate with the number of laser photons absorbed to reach the final state.
This can be modelled by a shift of $-\frac{\pi}{2}$ of the higher ATI peak as compared to the lower adjacent ATI peak.
Following similar steps as for RABBIT, we obtain for photoelectrons in the {\it up} direction
\begin{equation}
  \label{eq:zipp-prob}
  \begin{cases}
    |c_{\kbf}^{(\mathrm{SB})}|^2
    &= \xi^2\cos^2\left(\frac{\pi}{4} + \varphi \right) + \mathcal{O}(\xi^4) \\
    |c_\kbf^{(\mathrm{ATI})}|^2
    &= 1 - \xi^2\cos^2\left(\frac{\pi}{4} + \varphi\right) + \mathcal{O}(\xi^4),
  \end{cases}
\end{equation}
in excellent agreement with Fig.~\ref{fig:zipp}(b).
In Fig.~\ref{fig:zipp}~(c,d) we show a sideband and an ATI peak resolved over CEP and emission angle.
These angle-resolved ATI distributions are explained by the parity properties of the $2\omega$-ATI peaks: $f_{\mathbf{k}}^{(\mathrm{ATI})} \propto (-i)^{m} J_{m}$, where $m$ is the number of $2\omega$ photons absorbed.
Lastly, the relative phase shift between odd and even orders of HHG from an $\omega/2\omega$ field~\cite{LaurentPRL2012} can be interpreted analogously as an interaction phase on HHG process by the perturbative $2\omega$ field.
The associated interference effects have already been  discussed in Ref.~\cite{DahlstromJPB2011}.

In this Letter we have studied theoretically the interpretation of quantum dynamics in a range of recent attosecond experiments that rely on weak probe laser beams and introduced the concept of {\it interaction phases} that accumulate with each laser interaction.
We have found that RABBIT sidebands are exceptional, as they are {\it not} affected by such interaction phase shifts.
In contrast, experiments with an unbalanced number of photons may suffer from substantial CEP-shifts.
We have identified such effects in LAP by \textit{ab initio} simulation
of the checkerboard structure reported by Laurent et al.~\cite{LaurentPRL2012},
and of the FEL experiment reported by Maroju et al.~\cite{MarojuNature2020}.
Finally, we have shown that interaction phases can be used to interpret the action of a probe field on  strong-field processes, such as $\omega/2\omega$-HHG~\cite{LaurentPRL2012} and $2\omega/\omega$-ATI~\cite{zippOptica2014}.
In the former case, the interaction with a probe field leads to a phase shift between odd and even harmonics of $\pi/2$, while in the latter case, the accumulated interactions in $2\omega$-ATI lead to a CEP-shift of $\pi/4$, relative to a synchronized RABBIT experiment.
In addition to these interaction phases, which are derivable from the strong-field approximation, energy-dependent phases, such as scattering phases~\cite{starace_theory_1982} and continuum--continuum phases~\cite{KlunderPRL2011,DahlstromCP2013,FuchsOptica2020}, must also be considered to obtain attosecond precision in experiments.
This opens up a call for investigations of CEP effects on an {\it absolute} scale, relative to known reference experiments.

\subsection*{Acknowledgement}
JMD acknowledges support from the Swedish Research Council: 2018-03845, the Swedish Foundations' Starting Grant by the Olle Engkvist's Foundation: Grant No. 194-0734 and the Knut and Alice Wallenberg Foundation: Grant No. 2017.0104.
We thank Stefanos Carlström for fruitful discussions.


\begin{thebibliography}{10}

\bibitem{ZernePRA1997}
Raoul Zerne, Carlo Altucci, Marco Bellini, Mette~B. Gaarde, T.~W. H\"ansch,
  Anne L'Huillier, Claire Lyng\aa{}, and C.-G. Wahlstr\"om.
\newblock Phase-locked high-order harmonic sources.
\newblock {\em Phys. Rev. Lett.}, 79:1006--1009, Aug 1997.

\bibitem{XibinPRL2008}
Xibin Zhou, Robynne Lock, Wen Li, Nick Wagner, Margaret~M. Murnane, and
  Henry~C. Kapteyn.
\newblock Molecular recollision interferometry in high harmonic generation.
\newblock {\em Phys. Rev. Lett.}, 100:073902, Feb 2008.

\bibitem{SmirnovaNature2009}
Olga Smirnova, Yann Mairesse, Serguei Patchkovskii, Nirit Dudovich, David
  Villeneuve, Paul Corkum, and Misha~Yu. Ivanov.
\newblock High harmonic interferometry of multi-electron dynamics in molecules.
\newblock {\em Nature}, 460(7258):972--977, Aug 2009.

\bibitem{Kim2014NaturePhotonics}
Kyung~Taec Kim, D.~M. Villeneuve, and P.~B. Corkum.
\newblock Manipulating quantum paths for novel attosecond measurement methods.
\newblock {\em Nature Photonics}, 8(3):187--194, Mar 2014.

\bibitem{dudovichNaturePhysics2006}
N.~Dudovich, O.~Smirnova, J.~Levesque, Y.~Mairesse, M.~Yu Ivanov, D.~M.
  Villeneuve, and P.~B. Corkum.
\newblock Measuring and controlling the birth of attosecond {XUV} pulses.
\newblock {\em Nature Phys}, 2(11):781--786, November 2006.
\newblock Number: 11 Publisher: Nature Publishing Group.

\bibitem{DahlstromJPB2011}
J~M Dahlstr\"om, A~L\'Huillier, and J~Mauritsson.
\newblock Quantum mechanical approach to probing the birth of attosecond pulses
  using a two-colour field.
\newblock {\em Journal of Physics B: Atomic, Molecular and Optical Physics},
  44(9):095602, 2011.

\bibitem{ShafirNature2012}
Dror Shafir, Hadas Soifer, Barry~D. Bruner, Michal Dagan, Yann Mairesse,
  Serguei Patchkovskii, Misha~Yu. Ivanov, Olga Smirnova, and Nirit Dudovich.
\newblock Resolving the time when an electron exits a tunnelling barrier.
\newblock {\em Nature}, 485:343--346, May 2012.

\bibitem{PedatzurNaturePhysics2015}
O.~Pedatzur, G.~Orenstein, V.~Serbinenko, H.~Soifer, B.~D. Bruner, A.~J. Uzan,
  D.~S. Brambila, A.~ ~G. Harvey, L.~Torlina, F.~Morales, O.~Smirnova, and
  N.~Dudovich.
\newblock Attosecond tunnelling interferometry.
\newblock {\em Nature Physics}, 11(10):815--819, Oct 2015.

\bibitem{UzanNaturePhotonics2020}
Ayelet~J. Uzan, Hadas Soifer, Oren Pedatzur, Alex Clergerie, Sylvain Larroque,
  Barry~D. Bruner, Bernard Pons, Misha Ivanov, Olga Smirnova, and Nirit
  Dudovich.
\newblock Spatial molecular interferometry via multidimensional high-harmonic
  spectroscopy.
\newblock {\em Nature Photonics}, 14(3):188--194, Mar 2020.

\bibitem{VampaJPB2017}
G~Vampa and T~Brabec.
\newblock Merge of high harmonic generation from gases and solids and its
  implications for attosecond science.
\newblock {\em Journal of Physics B: Atomic, Molecular and Optical Physics},
  50(8):083001, mar 2017.

\bibitem{PaulScience2001}
P.~M. Paul, E.~S. Toma, P.~Breger, G.~Mullot, F.~Augé, Ph. Balcou, H.~G.
  Muller, and P.~Agostini.
\newblock Observation of a train of attosecond pulses from high harmonic
  generation.
\newblock {\em Science}, 292(5522):1689--1692, 2001.

\bibitem{KlunderPRL2011}
K.~Kl\"under, J.~M. Dahlstr\"om, M.~Gisselbrecht, T.~Fordell, M.~Swoboda,
  D.~Gu\'enot, P.~Johnsson, J.~Caillat, J.~Mauritsson, A.~Maquet, R.~Ta\"ieb,
  and A.~L'Huillier.
\newblock Probing single-photon ionization on the attosecond time scale.
\newblock {\em Phys. Rev. Lett.}, 106:143002, 2011.

\bibitem{LaurentPRL2012}
G.~Laurent, W.~Cao, H.~Li, Z.~Wang, I.~Ben-Itzhak, and C.~L. Cocke.
\newblock Attosecond control of orbital parity mix interferences and the
  relative phase of even and odd harmonics in an attosecond pulse train.
\newblock {\em Phys. Rev. Lett.}, 109:083001, 2012.

\bibitem{DahlstromCP2013}
J.M. Dahlstr\"om, D.~Gu\'enot, K.~Kl\"under, M.~Gisselbrecht, J.~Mauritsson,
  A.~L'Huillier, A.~Maquet, and R.~Ta\"ieb.
\newblock Theory of attosecond delays in laser-assisted photoionization.
\newblock {\em Chemical Physics}, 414(0):53 -- 64, 2013.

\bibitem{PazourekRevModPhys2015}
Renate Pazourek, Stefan Nagele, and Joachim Burgd\"orfer.
\newblock Attosecond chronoscopy of photoemission.
\newblock {\em Rev. Mod. Phys.}, 87:765--802, Aug 2015.

\bibitem{CaillatPRL2011}
J\'er\'emie Caillat, Alfred Maquet, Stefan Haessler, Baptiste Fabre, Thierry
  Ruchon, Pascal Sali\`eres, Yann Mairesse, and Richard Ta\"ieb.
\newblock Attosecond resolved electron release in two-color near-threshold
  photoionization of n2.
\newblock {\em Phys. Rev. Lett.}, 106:093002, 2011.

\bibitem{SwobodaPRL2010}
M.~Swoboda, T.~Fordell, K.~Kl\"under, J.~M. Dahlstr\"om, M.~Miranda, C.~Buth,
  K.~J. Schafer, J.~Mauritsson, A.~L'Huillier, and M.~Gisselbrecht.
\newblock Phase measurement of resonant two-photon ionization in helium.
\newblock {\em Phys. Rev. Lett.}, 104:103003, 2010.

\bibitem{JimenezPRL2014}
Alvaro Jimenez-Galan, Luca Argenti, and Fernando Martin.
\newblock Modulation of attosecond beating in resonant two-photon ionization.
\newblock {\em Phys. Rev. Lett.}, 113:263001, 2014.

\bibitem{KoturNatCom2016}
M.~Kotur, D.~Gu{\'e}not, {\'A}~Jim{\'e}nez-Gal{\'a}n, D.~Kroon, E.~W. Larsen,
  M.~Louisy, S.~Bengtsson, M.~Miranda, J.~Mauritsson, C.~L. Arnold, S.~E.
  Canton, T.~Gisselbrecht, M.~andCarette, J.~M. Dahlstr{\"o}m, E.~Lindroth,
  A.~Maquet, L.~Argenti, F.~Mart{\'i}n, and A.~L'Huillier.
\newblock Spectral phase measurement of a fano resonance using tunable
  attosecond pulses.
\newblock {\em Nature Communications}, 7:10566, 2016.

\bibitem{BarreauPRL2019}
Lou Barreau, C.~Leon~M. Petersson, Markus Klinker, Antoine Camper, Carlos
  Marante, Timothy Gorman, Dietrich Kiesewetter, Luca Argenti, Pierre Agostini,
  Jes\'us Gonz\'alez-V\'azquez, Pascal Sali\`eres, Louis~F. DiMauro, and
  Fernando Mart\'{\i}n.
\newblock Disentangling spectral phases of interfering autoionizing states from
  attosecond interferometric measurements.
\newblock {\em Phys. Rev. Lett.}, 122:253203, Jun 2019.

\bibitem{zippOptica2014}
Lucas~J. Zipp, Adi Natan, and Philip~H. Bucksbaum.
\newblock Probing electron delays in above-threshold ionization.
\newblock {\em Optica}, [1]([6]), November 2014.
\newblock Institution: SLAC National Accelerator Lab., Menlo Park, CA (United
  States) Publisher: Optical Society of America.

\bibitem{MarojuNature2020}
Praveen~Kumar Maroju, Cesare Grazioli, Michele~Di Fraia, Matteo Moioli, Dominik
  Ertel, Hamed Ahmadi, Oksana Plekan, Paola Finetti, Enrico Allaria, Luca
  Giannessi, Giovanni~De Ninno, Carlo Spezzani, Giuseppe Penco, Simone
  Spampinati, Alexander Demidovich, Miltcho~B. Danailov, Roberto Borghes,
  George Kourousias, Carlos Eduardo Sanches~Dos Reis, Fulvio Billé, Alberto~A.
  Lutman, Richard~J. Squibb, Raimund Feifel, Paolo Carpeggiani, Maurizio
  Reduzzi, Tommaso Mazza, Michael Meyer, Samuel Bengtsson, Neven Ibrakovic,
  Emma~Rose Simpson, Johan Mauritsson, Tamás Csizmadia, Mathieu Dumergue,
  Sergei Kühn, Harshitha~Nandiga Gopalakrishna, Daehyun You, Kiyoshi Ueda,
  Marie Labeye, Jens~Egebjerg Bækhøj, Kenneth~J. Schafer, Elena~V. Gryzlova,
  Alexei~N. Grum-Grzhimailo, Kevin~C. Prince, Carlo Callegari, and Giuseppe
  Sansone.
\newblock Attosecond pulse shaping using a seeded free-electron laser.
\newblock {\em Nature}, 578(7795):386--391, February 2020.

\bibitem{you_new_2020}
Daehyun You, Kiyoshi Ueda, Elena~V. Gryzlova, Alexei~N. Grum-Grzhimailo,
  Maria~M. Popova, Ekaterina~I. Staroselskaya, Oyunbileg Tugs, Yuki Orimo,
  Takeshi Sato, Kenichi~L. Ishikawa, Paolo~Antonio Carpeggiani, Tamás
  Csizmadia, Miklós Füle, Giuseppe Sansone, Praveen~Kumar Maroju, Alessandro
  D’Elia, Tommaso Mazza, Michael Meyer, Carlo Callegari, Michele Di~Fraia,
  Oksana Plekan, Robert Richter, Luca Giannessi, Enrico Allaria, Giovanni
  De~Ninno, Mauro Trovò, Laura Badano, Bruno Diviacco, Giulio Gaio, David
  Gauthier, Najmeh Mirian, Giuseppe Penco, Primož~Rebernik Ribič, Simone
  Spampinati, Carlo Spezzani, and Kevin~C. Prince.
\newblock New {Method} for {Measuring} {Angle}-{Resolved} {Phases} in
  {Photoemission}.
\newblock {\em Phys. Rev. X}, 10(3):031070, September 2020.
\newblock Publisher: American Physical Society.

\bibitem{rohringer_configuration-interaction-based_2006}
Nina Rohringer, Ariel Gordon, and Robin Santra.
\newblock Configuration-interaction-based time-dependent orbital approach for
  ab initio treatment of electronic dynamics in a strong optical laser field.
\newblock {\em Phys. Rev. A}, 74(4):043420, October 2006.

\bibitem{GreenmanPRA2010}
Loren Greenman, Phay~J. Ho, Stefan Pabst, Eugene Kamarchik, David~A. Mazziotti,
  and Robin Santra.
\newblock Implementation of the time-dependent configuration-interaction
  singles method for atomic strong-field processes.
\newblock {\em Phys. Rev. A}, 82:023406, 2010.

\bibitem{tao_photo-electron_2012}
Liang Tao and Armin Scrinzi.
\newblock Photo-electron momentum spectra from minimal volumes: the
  time-dependent surface flux method.
\newblock {\em New Journal of Physics}, 14:013021, January 2012.

\bibitem{keldysh_ionization_1965}
Leonid~V. Keldysh.
\newblock Ionization in the field of a strong electromagnetic wave.
\newblock {\em Sov. Phys. JETP}, 20(5), May 1965.

\bibitem{faisal_multiple_1973}
F.~H.~M. Faisal.
\newblock Multiple absorption of laser photons by atoms.
\newblock {\em J. Phys. B: At. Mol. Phys.}, 6(4):L89--L92, April 1973.
\newblock Publisher: IOP Publishing.

\bibitem{reiss_effect_1980}
Howard~R. Reiss.
\newblock Effect of an intense electromagnetic field on a weakly bound system.
\newblock {\em Phys. Rev. A}, 22(5):1786--1813, November 1980.
\newblock Publisher: American Physical Society.

\bibitem{MauritssonPRL2006}
J.~Mauritsson, P.~Johnsson, E.~Gustafsson, A.~L'Huillier, K.~J. Schafer, and
  M.~B. Gaarde.
\newblock Attosecond pulse trains generated using two color laser fields.
\newblock {\em Phys. Rev. Lett.}, 97:013001, 2006.

\bibitem{starace_theory_1982}
A.~F. Starace.
\newblock Theory of {Atomic} {Photoionization}.
\newblock {\em Handbuch der Physik}, 6:1, 1982.

\bibitem{kitzler_quantum_2002}
Markus Kitzler, Nenad Milosevic, Armin Scrinzi, Ferenc Krausz, and Thomas
  Brabec.
\newblock Quantum {Theory} of {Attosecond} {XUV} {Pulse} {Measurement} by
  {Laser} {Dressed} {Photoionization}.
\newblock {\em Phys. Rev. Lett.}, 88(17):173904, April 2002.
\newblock Publisher: American Physical Society.

\bibitem{IsingerScience2017}
M.~Isinger, R.~J. Squibb, D.~Busto, S.~Zhong, A.~Harth, D.~Kroon, S.~Nandi,
  C.~L. Arnold, M.~Miranda, J.~M. Dahlstr{\"o}m, E.~Lindroth, R.~Feifel,
  M.~Gisselbrecht, and A.~L{\textquoteright}Huillier.
\newblock Photoionization in the time and frequency domain.
\newblock {\em Science}, 358(6365):893--896, 2017.

\bibitem{FuchsOptica2020}
Jaco Fuchs, Nicolas Douguet, Stefan Donsa, Fernando Martin, Joachim
  Burgd\"{o}rfer, Luca Argenti, Laura Cattaneo, and Ursula Keller.
\newblock Time delays from one-photon transitions in the continuum.
\newblock {\em Optica}, 7(2):154--161, Feb 2020.

\end{thebibliography}
\end{document}